        \documentstyle[12pt]{article}
        
        \title{The Geometry of classical change of signature}
        \author{
        {\bf Mauro Carfora}$^1$ and {\bf George Ellis}$^{1,2}$ \\ ~\\
        {\em $^1$SISSA, Via Beirut 2-4,}\\
        {\em 34013 Trieste, Italy} \\~\\
        {\em $^2$Department of Applied Mathematics,} \\
        {\em University of Cape Town, 7700 Rondebosch,}\\
        {\em Republic of South Africa} }

\begin{document}
        \maketitle

        \begin{abstract}
        The proposal of the possibility of change of signature in quantum
        cosmology has led to the study of this phenomenon in classical
        general relativity theory, where there has been some controversy
        about what is and is not possible. We here present a new analysis
        of such a change of signature, based on previous studies of the
        initial value problem in general relativity. We emphasize that
        there are various continuity suppositions one can make at a
        classical change of signature, and consider more general
        assumptions than made up to now. We confirm that in general such
        a change can take place even when the second fundamental form of
        the surface of change does not vanish.
        \end{abstract}
        \newpage

        \section{Introduction}
        Following on recent developments in quantum cosmology [1-3], a subject
        of some interest is the possibility of a change of signature in a
        classical space-time [4-12]. We discuss here in depth the geometry
        associated with such a classical change of signature. The results
        obtained differ depending on what smoothness assumptions one makes. We
        look at the most general case, resulting from concentrating on the
        3-dimensional surface where the change of signature occurs, rather
        than on either the Lorentzian (hyperbolic) or Riemannian (positive
        definite) enveloping space (the latter is often referred to as
        Euclidean; however we prefer Riemannian, as `Euclidean' suggests that
        the space is flat) . \\

        In our approach we emphasize the initial value problem associated with
        signature change and the dynamical content of the theory, rather than
        regarding the problem as just a generalisation of the well-known
        Israel junction conditions [13]. There are more than junction
        conditions involved. In the case of the surface of a star, junction
        conditions are rather separated from the role of the initial value
        problem (because the surface is timelike). In the case of a change of
        signature, this must take place on a spacelike surface and so is
        essentially tied in to the nature of the initial value problem.
        Junction conditions play here a kinematical role, while the real
        dynamics of the change of signature are captured by the constraints
        associated with the field equations. This understanding underlies the
        approach we adopt. \\

        The first fundamental form must be continuous. The continuity of the
        the second fundamental form as seen from both sides, is only assumed
        up to the action of an infinitesimal diffeomorphism corresponding to a
        Lie derivative. This allows a kink in the geometry - not allowed in the
        more restrictive assumptions considered up to now. We insist that the
        constraints are valid for both enveloping metrics. Further junction
        conditions only arise if the matter is assumed to be smoothly behaved -
        which may not be required.\\

        These conditions thus generalise those considered by Ellis et al
        [5,6,11], which in turn are more general than those considered by
        Hayward et al [7,8,10] on the basis of their more restricted approach
        (placing more stringent restrictions on what is allowed). Our stand-
        point is that one can adopt any of these views - they are based on
        different philosophies of how one should approach junction conditions -
        or indeed one can question whether there should be any conditions other
        than a gluing condition, such as is adopted here. \\

        We avoid use of specific coordinate systems, as well as use of
        abstract notation such as is employed by Hayward [7]. Rather we follow
        the notation of Hawking and Ellis [14] and of Fisher and Marsden [15].

        \section{Approach Taken}
        We let ${\cal S}$ denote a compact oriented three-manifold, and let
        \begin{equation}
        \Theta: {\cal S} \rightarrow (M^{(4)},{\bf g}) \equiv M
        \end{equation}
        be an embedding of ${\cal S}$ in a Lorentzian manifold
        $(M^{(4)},{\bf g})$ such that the imbedded manifold $\Theta({\cal S})$
        is space-like, that is the pull-back
        \begin{equation}
        \Theta^*({\bf g}) \equiv {\bf h}
        \end{equation}
        is a Riemannian metric on ${\cal S}$. \\

        Similarly we define
        \begin{equation}
        \hat{\Theta}: {\cal S} \rightarrow (M^{(4)},\hat{{\bf g}}) \equiv
        \hat{M}
        \end{equation}
        as an embedding of ${\cal S}$ in the {\it same} 4-dimensional manifold,
        $M^{(4)}$, but now endowed with a Riemannian metric $\hat{{\bf g}}$,
        {\it viz.},
        $(M^{(4)},\hat{{\bf g}})$. \\

        Our strategy is to think of the metrics ${\bf g}$ and ${\bf \hat{g}}$
        as living on the same portion of manifold, and in order to avoid
        misunderstandings, we wish to stress that $M$ and $\hat{M}$ are
        just a shorthand notation for the same underlying four-manifold
        $M^{(4)}$ with different metrics, with ${\bf g}$ Lorentzian,
        whereas ${\hat{\bf g}}$ is Riemannian. As we are not concerned
        with global problems we may restrict ourselves to a a tubular
        neighborhood of $\Theta({\cal S})$ (containing also
        $\hat{\Theta}({\cal S})$). For the moment ${\bf g}$ and ${\bf
        \hat{g}}$ are arbitrary. This coexistence of both Riemannian and
        Lorentzian metrics on the same region of the manifold will in
        our opinion avoid a lot of problems when thinking of the geometry
        involved. \\

        We are going to identify - modulo the action of the diffeomorphisms-
        the Lorentzian and Riemannian geometry along a common imbedded
        space-like hypersurface, determined by the constraints associated with
        the Einstein equations.

        \section{Geometry}
        In order to define the variables of interest, we need to characterise
        the foliations employed and the related lapse and shift in both the
        Riemannian and Lorentzian cases.\\

        Let $E^\infty({\cal S},\hat{M})$  and $E^\infty({\cal S},M)$ denote
        the sets of all spacelike imbeddings of ${\cal S}$ in $\hat{M}$ and
        $M$ respectively. \\

        Suppose we have a curve in each of these imbedding spaces: namely a
        one-parameter $(\lambda)$ family of spacelike imbeddings of ${\cal S}$
        into $M$, and a similar one-parameter $(\lambda)$ family of imbeddings
        of ${\cal S}$ into $\hat{M}$. Explicitly,
        ${\Theta}_{\lambda}\colon{\cal S}\times{I}\to{M}$ and
        $\hat{\Theta}_{\lambda}\colon{\cal S}\times{I}\to\hat{M}$, where
        $I\equiv{(-\epsilon,\epsilon)}$ for a suitably small $\epsilon>0$.
         This family of imbeddings defines a
        corresponding one-parameter family of vector fields
        $X^{(4)}_\lambda\colon{\cal S}\to{TM^{(4)}}$
        and $\hat{X}^{(4)}_\lambda\colon{\cal S}\to{T\hat{M}^{(4)}} $ by
        \begin{equation}
        {d\Theta_\lambda \over d\lambda}(p) = X^{(4)}_\lambda(\Theta_\lambda
        (p))
        \end{equation}
        and
        \begin{equation}
        {d\hat{\Theta}_\lambda \over d\lambda}(p) =
        \hat{X}^{(4)}_\lambda(\hat{\Theta}_\lambda (p))
        \end{equation}
        as $p$ varies over ${\cal S}$. \\

        In order to simplify the notation a bit, we shall denote them simply
        by $X_\lambda$ and $\hat{X}_\lambda$. Roughly speaking, either in $M$
        or in $\hat{M}$ these vectors connect the point $\Theta_\lambda(p)$
        with $\Theta_{\lambda+d\lambda}(p)$ (and similarly for
        $\hat{\Theta}$); namely the images of a given point $p$ in ${\cal S}$
        under two infinitesimally near imbeddings. \\

        If $n$ and $\hat{n}$ respectively denote the forward-pointing unit
        normals to $\Theta({\cal S})$ and $\hat{\Theta}({\cal S})$ (so $n^a
        n^b g_{ab} = -1$; $\hat{n^a} \hat{n}^b \hat{g}_{ab} = +1$), we can
        as usual decompose the vector fields $X$ and $\hat{X}$ into their
        normal and tangential components:
        \begin{equation}
        X_\lambda = N_\lambda \hat{n} + \beta_\lambda
        \end{equation}
        \begin{equation}
        \hat{X}_\lambda = \hat{N}_\lambda \hat{n} + \hat{\beta}_\lambda
        \end{equation}
        which define the corresponding family of lapse functions on
        ${\cal S}$, {\it i.e.},
        $N_{\lambda}\colon{\cal S}\to{\bf R}$ and a corresponding family of
        shift vector fields again on ${\cal S}$, namely
        ${\beta}_{\lambda}\colon{\cal S}\to{T{\cal S}}$.
        We wish to stress the fact (slightly obscured by our simplified nota-
        tion) that the family of lapse functions $N_\lambda$ are defined on the
        abstract manifold ${\cal S}$, and similarly the family of shift vector
        fields $\beta_\lambda$ are defined over ${\cal S}$; similarly for the
        lapse $\hat{N}_\lambda$ and shift $\hat{\beta}_\lambda$. Here ``the
        lapse and the shift are seen in their proper geometric roles - describ-
        ing the hypersurface deformations in the enveloping geometries - rather
        than as pieces of the metric'' (Isenberg and Nester [16]). \\

        The metric interpretation comes about for instance if we use the maps
        \begin{equation}
        F: I \times {\cal S} \rightarrow M
        \end{equation}
        defined by
        \begin{equation}
        (\lambda,p) \mapsto \Theta_\lambda(p)
        \end{equation}
        as a diffeomorphism of $I \times {\cal S}$ onto a
        tubular neighbourhood of $\Theta_0({\cal S})$. We can then pull back
        the metric $g$ onto $I \times {\cal S}$ and get the usual expression
        \begin{equation}
        (F^* g)_{\alpha\beta} dx^\alpha dx^\beta = - (N^2_\lambda - \beta_i
        \beta^i)d\lambda^2 + 2 \beta_i dx^i d\lambda + h_{ij} dx^i dx^j
        \end{equation}
        where indices $\alpha$ and $\beta$ run from 1 to 4, $i$ and $j$ run
        from 1 to  3, $\{x^i\}$ are local coordinates on ${\cal S}$, and
        $h_{ij}$ is the $\lambda-$dependent one-parameter family of metrics on
        ${\cal S}$.  A similar analysis holds for $\hat{F}$, leading to
        \begin{equation}
        (\hat{F}^* \hat{g})_{\alpha\beta} dy^\alpha dy^\beta = +
        (\hat{N}^2_\lambda + \hat{\beta}_i \hat{\beta}^i)d\lambda^2 + 2
        \hat{\beta}_i dy^i d\lambda + \hat{h}_{ij} dy^i dy^j
        \end{equation}
        with the obvious meaning of the symbols. \\

        There are a number of general comments that should be made at this
        stage. In particular, we wish to caution the reader to not confuse the
        abstract manifold ${\cal S}\times{I}$ with its images
        ${\Theta}_{\lambda}({\cal S})$ and
        $\hat{\Theta}_{\lambda}({\cal S})$ in $M=(M^{(4)},{\bf g})$
        and $\hat{M}=(M^{(4)},\hat{\bf g})$, respectively. Typically, when
        dealing with the initial value problem, one is accustomed to do so for
        obvious reasons, and this identification is usually harmless. However
        making clear the distinction is more than a technical convenience here.
        By identifying ${\cal S}\times{I}$ with ${\Theta}_{\lambda}({\cal S})$
        and $\hat{\Theta}_{\lambda}({\cal S})$ one is lead to an incorrect
        interpretation of the vector field $\partial/\partial{\lambda}$, which
        is defined on ${\cal S}\times{I}$, in terms of which the initial value
        formalism is phrased. Observe that the parameter $\lambda$ is the
        natural label for all the fields ${\bf h}_{\lambda}$, ${\hat {\bf
        h}}_{\lambda}$, $N_{\lambda}$, ${\hat N}_{\lambda}$,
        ${\beta}_{\lambda}$, ${\hat{\beta}}_{\lambda}$, and the extrinsic
        curvatures (defined below), if they are referred either to the Lorent-
        zian or to the  Riemannian case. This is a rather obvious statement
        when things are correctly seen, as they should be, on ${\cal
        S}\times{I}$. It is not an obvious statement at all if we identify
        ${\cal S}\times{I}$ with its images under ${\Theta}_{\lambda}$ and
        $\hat{\Theta}_{\lambda}$. In this case, since the foliations
        ${\Theta}_{\lambda}({\cal S)}$ and $\hat{\Theta}_{\lambda}{\cal S})$
        are different,  and with different deformation vectors $X_{\lambda}$
        and ${\hat {X}}_{\lambda}$, one is incorrectly led to believe that
        these deformation vectors must be tangent to different deformation
        coordinates, namely $X_{\lambda} =\partial/\partial{\lambda}$ and
        ${\hat {X}}_{\lambda}=\partial/\partial{\omega}$, for some other defor-
        mation parameter ${\omega}$. As stressed before, this is usually harm-
        less in standard situations where one has just one enveloping space-
        time, but it is fatal here where the enveloping geometries are two and
        quite distinct.\par

         The source of the error, in proceeding as above, lies in the fact that
        one is identifying vectors living on different spaces, since the family
        of vector fields $\partial/\partial{\lambda}$ is defined on ${\cal S}$,
        while the deformation vectors $X_{\lambda}$ and $\hat{X}_{\lambda}$ are
        defined on $M$ and $\hat{M}$, respectively. If these two latter are
        different, their intuitive identification with vectors tangent to a
        deformation coordinate, ({\it i.e.}, with
        $\partial/\partial{\lambda}$), is problematic and very confusing.\par

        One must clearly separate the role of the vector tangent to the
        deformation coordinate,
        which is  $\partial/\partial{\lambda}$, and which is defined on
        ${\cal S}$, from the vectors
        $X_{\lambda}$ and $\hat{X}_{\lambda}$  which are respectively
        associated to the imbeddings ${\Theta}_{\lambda}$ and
        $\hat{\Theta}_{\lambda}$, (these vector fields can be
        thought of as the vector fields
        covering the two distinct family of imbeddings
        ${\Theta}_{\lambda}$ and $\hat{\Theta}_{\lambda}$ of ${\cal S}$).\\

        It is our strategy to address the geometry of signature change exclu-
        sively in terms of quantities defined on ${\cal S}$ and this should be
        clearly kept in mind when deciding which quantities should be
        continuous through a surface of signature change. For instance it would
        be very unnatural from our viewpoint to assume the continuity
        of the unit normals, for these quantities live in the embedding
        spacetimes $M$ and $\hat{M}$, and this is something that an
        observer living in ${\cal S}$ does not know {\it a priori}.
         It is much more natural for him to assume the continuity of
        the vector $\partial/\partial{\lambda}$ and
        of the lapse function and of the shift vector fields, since all such
        quantities are well defined on ${\cal S}$ and they provide him
        the complete kinematical framework for describing -- from his
        standpoint -- the deformations of ${\cal S}$ which may be
        compatible with a Riemannian geometry on one side and with
        a Lorentzian geometry on the other.\\

        With these general remarks out of the way, we recall that in order
        to describe the imbeddings $\Theta$ and $\hat{\Theta}$, besides
        introducing the 3-metrics $h$ and $\hat{h}$ we must also introduce,
        on ${\cal S}$, two symmetric tensor fields $K$ and $\hat{K}$ to be
        interpreted as  the second fundamental forms of
        $\Theta_\lambda({\cal S})$ and $\hat{\Theta}_\lambda({\cal S})$
        respectively. In our notation, they are defined, at the generic point
        $x\in{\cal S}$, and for any pair of vectors ${\bf u}$ and ${\bf v}$ in
        $T_x{\cal S}$ by
        \begin{equation}
        K_x({\bf u},{\bf v}) =<T_x{\Theta}\circ{\bf u}|
        {\nabla}^{(4)}(T_x{\Theta}\circ{\bf v}){\bf n}>_{g}({\Theta}(x))
        \end{equation}
        where ${\nabla}^{(4)}$ denotes the covariant derivative operator in
        $M$, the brackets $<\cdot|\cdot>_{g}({\Theta}(x))$ stand for the inner
        product in the Lorentzian metric ${\bf g}$ evaluated at the point
        ${\Theta}(x)\in{M}$, and $T_x{\Theta}$ stands for the tangential
        mapping, at $x\in {\cal S}$, associated to the embedding ${\Theta}$.

        Similarly, and with an obvious meaning of the symbols,
        \begin{equation}
        \hat{K}_x({\bf u},{\bf v}) = <T_x{\hat{\Theta}}\circ{\bf u}|
        {\hat{\nabla}}^{(4)}
        (T_x{\hat{\Theta}}\circ{\bf v}){\bf n}>_{\hat{g}}
        ({\hat{\Theta}}(x))
        \end{equation}

        For each given value of $\lambda$ the fields $(h,K)$ and $(\hat{h},
        \hat{K})$ cannot be arbitrarily prescribed. From the Gauss-Codazzi
        relation, one gets that such fields must satisfy four compatibility
        conditions, namely in the Riemannian case
        \begin{equation}
        R(\hat{h}) - (\hat{K}^{dc} \hat{h}_{dc})^2 + \hat{K}^{ab}
        \hat{K}^{cd} \hat{h}_{ac} \hat{h}_{bd} = -
        2 \Theta^*(G_{\mu\nu}\hat{n}^{\mu}\hat{n}^{\nu})
        \end{equation}
        where $\hat{G}_{\mu\nu}$ is the Einstein tensor of $\hat{g}$
        and
        \begin{equation}
        \hat{D}_a \hat{K}^{ac} \hat{h}_{cb} - \hat{D}_b \hat{K}^{cd}
        \hat{h}_{cd} =
        \hat{\Theta}^* [R_{\mu\nu}(\hat{g}) \hat{n}^{\nu}{\perp}^{\mu}_b]
        \end{equation}
        where $\hat{D}$ is the covariant derivative in $({\cal S},\hat{h})$
        and $R_{\mu\nu}$ is the Ricci tensor of the metric $\hat{g}$.\\

        In the Lorentzian case, we obtain
        \begin{equation}
        R(h) + ( K^{dc}  h_{dc})^2 -  K^{ab}
         K^{cd}  h_{ac}  h_{bd} = 2 \Theta^*(G_{\mu\nu}n^{\mu}n^{\nu})
        \end{equation}
        where $ G_{\mu\nu}$ is the Einstein tensor of $g$ and
        \begin{equation}
        D_a K^{ac} h_{cb} - D_b K^{cd} h_{cd} =
        \Theta^* [R_{\mu\nu}(g)n^{\nu}{\perp}^{\mu}_b] \, .
        \end{equation}

        \section{Change of Signature}
        Now we are ready to  discuss the possibility of change of signature
        through a regular hypersurface. Till now the embedded hypersurfaces
        $\Theta_\lambda({\cal S})$ and $\hat{\Theta}_\lambda({\cal S})$ were
        kept distinct. The basic condition we need in order to be able to
        speak of a signature change is to choose one of the $\Theta_\lambda
        ({\cal S})$ to `coincide' with one of the manifolds of the family
        $\hat{\Theta}_\lambda({\cal S})$. \\

        Asking directly, as often is implicitly done, that for a given
        range of $\lambda$, say $-\epsilon<\lambda<\epsilon$,
        ${\Theta}_\lambda({\cal S})\equiv \hat{\Theta}_\lambda({\cal S})$,
        is too restrictive. And this is partially the reason for having
        unnecessary
        stringent constraints on the second fundamental form on the
        hypersurface of signature change. It is  more natural to assume, at
leas
   t
        a priori, that
        the identification between ${\Theta}_\lambda({\cal S})$ and
        $\hat{\Theta}_\lambda({\cal S})$, $-\epsilon<\lambda<\epsilon$,
        occurs modulo the action of diffeomorphisms of the manifold
        ${\cal S}$. More particularly, we consider a $\lambda$ dependent
        family of diffeomorphisms ${\phi}_{\lambda}\colon{\cal S}\to{\cal S}$,
        smoothly varying as $-\epsilon<\lambda<\epsilon$, and such that
        for a given value of $\lambda$, say $\lambda=0$,
        \begin{equation}
        \hat{\Theta}_0(p)={\Theta}_0(p), \forall p\in{\cal S}
        \end{equation}
        namely, it is only required that ${\phi}_{\lambda}=id_{\cal S}$ for
        $\lambda=0$. The strategy will be to use these diffeomorphisms to
        glue the bottom (Riemannian) region with the top (Lorentzian) region.
        This will mean - remembering that there are two metrics on
        ${\cal S}\times{I}$
         - we designate the metric $\hat{g}$ as the physical metric in
        the lower region ${\cal S}\times(0,-\beta)$  and the metric $g$ as the
        physical metric in the upper region ${\cal S}\times(\beta,0)$.
        On the zero section, ${\cal S}\times\{0\}$, of ${\cal S}\times{I}$,
        the  constraints associated to the Lorentzian and to the Riemannian
        imbedding must be simultaneously satisfied. \\

        It is clear that as far as the three-metrics $h$ and $\hat{h}$ are
        concerned, the action of the one-parameter group of diffeomorphisms
        ${\phi}_{\lambda}$ is simply that of having

        \begin{equation}
        \hat{h} = {\phi}_{\lambda}^* h
        \end{equation}
        for $-\epsilon<\lambda<\epsilon$, and in particular,
        $\hat{h} = h$ for $\lambda=0$.\\

        The situation is less dull as far as concerns the tensor
        fields $K$ and $\hat{K}$ yielding the second fundamental
        forms. In order to see how the action of ${\phi}_{\lambda}$
        relates $K$ and $\hat{K}$ on ${\cal S}$ let us write the
        explicit expressions of $K$ and $\hat{K}$ in terms of
        the three-metrics $h$, $\hat{h}$, and of the vector
        field (defined over ${\cal S} \times I$),
        $\frac{\partial}{\partial\lambda}$. We get

        \begin{equation}
        K_{ij}=N_{\lambda}^{-1}[\frac{\partial}{\partial{\lambda}}h_{ij}-
        L_{{\beta}_{\lambda}}h_{ij}]
        \end{equation}
        and similarly

        \begin{equation}
        \hat{K}_{ij}=\hat{N}_{\lambda}^{-1}
        [\frac{\partial}{\partial{\lambda}}\hat{h}_{ij}-
        L_{\hat{\beta}_{\lambda}}\hat{h}_{ij}]
        \end{equation}
        where $L_{\cdot}$ denotes Lie differentiation along the vector
        field indicated. \\

        For $-\epsilon<\lambda<\epsilon$, we have
        $\hat{h}_{ij}=({\phi}^*_{\lambda}h)_{ij}$ thus

        \begin{equation}
        \hat{K}_{ij}=\hat{N}_{\lambda}^{-1}
        [\frac{\partial}{\partial{\lambda}}({\phi}^*_{\lambda}h)_{ij}-
        L_{\hat{\beta}_{\lambda}}({\phi}^*_{\lambda}h)_{ij}]
        \end{equation}

        A direct computation shows (see {\it e.g.}, DeTurck [17]),

        \begin{equation}
        \frac{\partial}{\partial{\lambda}}[({\phi}^*_{\lambda}h)_{ij}(p)]=
        {\phi}^*_{\lambda}
        [\frac{\partial}{\partial{\lambda}}h_{ij}({\phi}_{\lambda}(p))]+
        {\phi}^*_{\lambda}
        [L_{{v}_{\lambda}}h_{ij}({\phi}_{\lambda}(p))]
        \end{equation}
        where the vector field $v_{\lambda}$ is the generator of the
        one-parameter group of diffeomorphisms ${\phi}_{\lambda}$
        according to

        \begin{equation}
        \frac{\partial}{\partial{\lambda}}{\phi}_{\lambda}(p)=
        v_{\lambda}(\lambda,{\phi}_{\lambda}(p))
        \end{equation}
        with the initial condition
        ${\phi}_{\lambda}|_{\lambda=0}=id_{\cal S}$.\\

        Thus
        \begin{equation}
        \hat{K}_{ij}=\hat{N}_{\lambda}^{-1}
        {\phi}^*_{\lambda}[
        \frac{\partial}{\partial{\lambda}}h_{ij}({\phi}_{\lambda}(p))+
        L_{{v}_{\lambda}}h_{ij}({\phi}_{\lambda}(p))-
        L_{\hat{\beta}_{\lambda}}h_{ij}({\phi}_{\lambda}(p))]
        \end{equation}
        In particular, for $\lambda=0$, we get

        \begin{equation}
        \hat{K}_{ij}=\hat{N}_{\lambda}^{-1}[
        \frac{\partial}{\partial{\lambda}}h_{ij}+
        L_{{v}_{\lambda}}h_{ij}-
        L_{\hat{\beta}_{\lambda}}h_{ij}]
        \end{equation}
        which shows that if, as argued in the previous paragraph, we
        assume continuity of the lapse and the
        shift for $\lambda=0$:

        \begin{equation}
        \hat{N}_{\lambda}=N_{\lambda},
        ~~~\hat{\beta}_{\lambda}={\beta}_{\lambda},
        \end{equation}
        and assuming also continuity of
        $\frac{\partial}{\partial{\lambda}}h_{ij}$, then

        \begin{equation}
        \hat{K}_{ij}=K_{ij}+ N_{\lambda}^{-1}L_{v_{\lambda}}h_{ij}
        \end{equation}
        (a similar relation holds for any $-\epsilon<\lambda<\epsilon$
        provided that we act by ${\phi}^*_{\lambda}$). Thus, on
        the hypersurface ${\Theta}_0({\cal S})=\hat{\Theta}_0({\cal S})$
        where we seek a change of signature, we may assume that the
        corresponding second fundamental forms coincide only up
        the Lie derivative term
        $N_{\lambda}^{-1}L_{v_{\lambda}}h^{ij}$. \\

        We wish to stress that by forcing ${\phi}_{\lambda}$ to be the
        identity for all $\lambda$, one may obviously achieve equality between
        the second fundamental forms on the transition hypersurface. But
        fixing {\it a priori} the three degrees of freedom (per space
        point) associated with ${\phi}_{\lambda}$ will be a very bad
        investment when dealing with the constraints. \\

        One may also argue that equation (26) is equally compatible with having
        continuity of the second fundamental form, provided that one allows
        for a discontinuous shift vector field, namely
        ${\beta}_{\lambda}=\hat{\beta}_{\lambda}-v_{\lambda}$. Further
impositio
   n
        of the continuity of the shift would then yield $v_{\lambda}=0$, and
        the former case of freezing the diffeomorphism group is then recovered.
        All this is actually related to what one considers standard junction
        conditions in the setting of signature changes. In ordinary situations,
        such conditions require the continuity of the four-metric and of
        the second fundamental form. But whether or not a such conditions
        can be extended at face value to the case of surfaces of signature
        change is a very delicate issue. Continuity of the four-metric
        leads to vanishing of the lapse function, which is quite disturbing.
        Moreover, the tensor fields $K$ and $\hat{K}$, when interpreted as
        second fundamental forms, are to be thought of as defined in terms of
        a unit normal (to the surface of signature change) whose norm
        changes sign at the junction. Thus it is not obvious at all that
        the continuity of the second fundamental form is a {\it natural}
        requirement in the case of surfaces of signature change.\\

        In this respect, it is often argued that the correct answer must
        come from the field equations. More precisely, one should impose
        the validity of the field equations everywhere, in particular on the
        surface of signature change. This point of view is apparently
        reasonable and interesting, but implies very severe constraints on the
        resulting solutions. We offer here an alternative point of view, namely
        we do not force the validity of the full four dimensional field
        equations on the surface of signature change, but rather we
        concentrate on the validity of that part of the field equations which
        is really {\it intrinsic} to the surface of signature change,
        namely we impose the consistency among the four constraints associated
        with the field equations. In our view, this is a minimal necessary
        requirement, the basic one. Further restrictions can come only if one
        has some input from the matter fields present, in particular on how
        they behave on the surface of signature change; and that is a matter
        for debate.\\

        We wish also to stress the following point. From the point of view of
        analysis and physics, partial differential equations of mixed type,
        where the type  (elliptic, hyperbolic, or parabolic) of the equation is
        a function of position, are rather familiar. The added difficulty here,
        in considering surfaces of signature change, lies exactly in the
        diffeomorphism invariance of the theory. By considering the full
        field equations at once everywhere, one is behaving as if there exists
        a general theory of boundary value problems independent of the type of
        the  equation, which is very bold, to say the least. Even in the
        simplest cases in hydrodynamics, such a theory is very delicate, and
        general results exist only for equations of special types. The
        situation becomes hopeless in a general relativity setting. Indeed,
        Einstein's equations in the Riemannian regime are a strongly
        overdetermined elliptic system (owing to diffeomorphism invariance),
        and the problem of finding a metric with a preassigned Einstein or
        Ricci tensor is often obstructed even at an infinitesimal level, ({\it
        i.e.}, there are even obstructions  to finding a metric, around a given
        point, with prescribed Ricci tensor, see [17]). The situation changes
        drastically in the Lorentzian regime. Thus it is fair to say that the
        study of mixed type Einstein equations is a completely open problem. It
        follows that forcing the validity of the field equations everywhere,
        in the case of a surface of signature change, is a formal procedure not
        really justified from an existing theory, and to which one should give
        the same {\it interlocutory} status as other proposals.
        In our approach, restricting attention to the constraints forced
        on the surface of signature change, one is considering what kind
        of initial data is compatible with a signature change in terms of
        partial differential equations which do {\it not} change type on
        the surface of signature change. Furthermore, these contain the
        essential dynamical equations of the theory (for example in the
        Robertson-Walker case, they include the Friedmann equation),
        which lead to the Wheeler-de Witt equation which underlies quan-
        tum cosmology.\\

        As a final remark, notice that at first reading one may think there is
        a surface layer present in our formalism because of the allowed
        discontinuity of the second fundamental form. However,  there is no
        variance with the essence of the junction conditions of Israel [13],
        since we are assuming the continuity of the proper dynamical variables,
        which are $\frac{\partial} {\partial {\lambda}} h_{ij}$. These
        conditions are usually written down in terms of adapted coordinates
        such that the second fundamental form is the time derivative, and so do
        not allow for the action of a diffeomorphism which is responsible for
        the Lie derivative terms. Actually, in the geometrical setting
        discussed here, as stressed above, they should not be taken at face
        value, since the general remarks discussed in the previous paragraph
        apply also here. In our setting, the proper variables to match are the
        lapse $N_{\lambda}$, the shift ${\beta}_{\lambda}$, the three-metric
        ${\bf h}_{\lambda}$ and  its derivative $\frac{\partial} {\partial
        {\lambda}}{\bf h}$ -- as we have done, and no surface layer is present
        as is clearly shown by imposing the constraints.

        \subsection{Constraints}
         The constraints, both in their Lorentzian and Riemannian version,
        must hold for $\lambda=0$.\\

        Let us start from the momentum (or divergence) constraint.
        We assume that on $M$ both $\hat{g}$ and $g$ satisfy the
        corresponding form of Einstein field equations, the Riemannian
        form for the former, the standard Lorentzian form for the latter. Thus
        in the Riemannian case

        \begin{equation}
        \hat{R}_{\alpha\beta}=\hat{T}_{\alpha\beta}-
        \frac{1}{2}\hat{g}_{\alpha\beta}
        \hat{g}^{\gamma\delta}\hat{T}_{\gamma\delta}
        \end{equation}
        where $\hat{T}_{\alpha\beta}$ are the components of the Riemannian
        energy-momentum tensor. Relative to the slicing
        $\hat{\Theta}_{\lambda}({\cal S})$
        we shall write

        \begin{equation}
        \hat{T}_{\alpha\beta}=\hat{\mu}\hat{n}_{\alpha}\hat{n}_{\beta}+
        \hat{j}_{\alpha}\hat{n}_{\beta}+
        \hat{j}_{\beta}\hat{n}_{\alpha}+\hat{s}_{\alpha\beta}
        \end{equation}
        where $\hat{\mu}$, $\hat{j}_{\alpha}$, and
        $\hat{s}_{\alpha\beta}$ respectively
        are the normal-normal, normal-tangential, and tangential-tangential
        projections of $\hat{T}_{\alpha\beta}$ with respect to
        $\hat{\Theta}_{\lambda}({\cal S})$. In the Lorentzian case, we
        shall similarly write

        \begin{equation}
        {R}_{\alpha\beta}={T}_{\alpha\beta}-
        \frac{1}{2}{g}_{\alpha\beta}
        {g}^{\gamma\delta}{T}_{\gamma\delta}
        \end{equation}
        where ${T}_{\alpha\beta}$ are the components of the energy-momentum
        tensor which, relative to the slicing
        ${\Theta}_{\lambda}({\cal S})$, can be decomposed according to

        \begin{equation}
        {T}_{\alpha\beta}={\mu}{n}_{\alpha}{n}_{\beta}+
        {j}_{\alpha}{n}_{\beta}+
        {j}_{\beta}{n}_{\alpha}+{s}_{\alpha\beta}
        \end{equation}
        where ${\mu}$, ${j}_{\alpha}$, and ${s}_{\alpha\beta}$ respectively
        are the relative density of mass-energy, the relative density of
        momentum, and the relative spatial stress tensor
         with respect to ${\Theta}_{\lambda}({\cal S})$.\\

        In general, there is no a priori need to assume that
        for $\lambda=0$ the {\it matter} variables are continuous. From a
        phenomenological point of view, there is no obvious evidence that one
        should assume continuity of the stress tensor components at the change
        of signature, although one might make that assumption if given no
        further information. On the other hand, if one has a more fundamental
        description of the stress tensor, for example as arising from a scalar
        field, one can work out the continuity properties of the stress tensor
        components from that description. This was done in [5,6] for the case
        of a classical scalar field. Then the obvious continuity conditions
        are that the fundamental variables associated with the more
        fundamental description are continuous, and satisfy whatever
        requirements there may be to give a good set of initial data for the
        matter field equations on either side of the signature change
        surface.\\

        In general this will result in discontinuous stress tensor
        components. This is not unreasonable in view of the fact that the
        usual conservation laws for energy and momentum break down at a change
        of signature surface [12]. The fundamental underlying point is that it
        is difficult to understand physics in the positive definite region,
        indeed classical physics in the usual sense will not exist there
        (although quantum physics will be fine!). Thus one must be open-minded
        as to what conditions should be imposed on `matter' in the positive
        definite regime, in a classical discussion of signature change. \\

        Without making specific assumptions, the momentum constraint forced on
        ${\cal S}\times\{0\}$ by the Riemannian side is

        \begin{equation}
        \hat{D}^a\hat{K}_{ab}-\hat{D}_b\hat{k}=\hat{j}^b
        \end{equation}
        where $\hat{k}\equiv \hat{h}^{cd}\hat{K}_{cd}$ is the
        rate of volume expansion (the trace of the second fundamental
        form). Since, for $\lambda=0$,
        $\hat{D}^a=D^a$ and $\hat{K}_{ab}=K_{ab}
        +N^{-1}L_vh_{ab}$ we get

        \begin{equation}

D^aK_{ab}+D^a(N^{-1}L_vh_{ab})-D_bk-D_b(h^{cd}N^{-1}L_vh_{cd})=\hat{j}_b
        \end{equation}
        (where as above $k\equiv h^{cd}K_{cd}$). But the momentum constraint
        forced on ${\cal S}\times\{0\}$ by the Lorentzian side implies that

        \begin{equation}
        {D}^a{K}_{ab}-{D}_b{k}=-{j}^b
        \end{equation}
        which, when introduced in the previous expression, yields

        \begin{equation}
        D^a(N^{-1}L_vh_{ab})- D_b(h^{cd}N^{-1}L_vh_{cd})=j_b+\hat{j}_b
        \end{equation}

        Given $h_{ab}$, the lapse function $N$, and the
        momentum densities $j_b$, $\hat{j}_b$ the above is a
        system of partial differential equations determining
        the vector field $v$ which generates the gluing one-parameter
        group of diffeomorphisms ${\phi}_{\lambda}$ in the neighbourhood
        of $\lambda=0$. Notice however that this system is elliptic
        ({\it i.e.}, (36) can be actually inverted) only if the vector
        field $v$ is divergence-free, $D^av_a=0$. This further requirement
        implies that  $k$, the trace of the second fundamental form, is
        continuous through the surface of signature change $S$, namely
        \begin{equation}
        k=\hat{k}
        \end{equation}
        This result is quite satisfactory since  in an
        initial value approach, the rate of volume expansion is to be
        considered as a kinematical variable selecting the family of
        hypersurfaces
        along which we are following the dynamics of the gravitational
        field.

        \vskip 0.5 cm
        Next we can impose that both the Riemannian and the Lorentzian
        version of the Hamiltonian constraint hold for ${\cal S}\times\{0\}$.
        This yields

        \begin{equation}
        2{\mu}+2\hat{\mu}+\frac{2}{3}k^2+\\

h^{ac}h^{bd}(\tilde{K}_{ab}+N^{-1}L_vh_{ab})(\tilde{K}_{cd}+N^{-1}L_vh_{
   cd})+
        \tilde{K}^{ab}\tilde{K}_{ab}=0
        \end{equation}
        where $\tilde{K}_{ab}$ denotes the trace-free part of
        $K_{ab}$. \par
        If we assume that $\hat{\mu}\geq 0$, then the above condition,
        being the sum of algebraically independent non-negative terms,
        is only compatible with the vanishing of each summand. Thus, in this
cas
   e
        from $\hat{\mu}\geq 0$ we actually get
        $\hat{\mu}=0$, $\mu=0$, $k=0$, $\tilde{K}_{ab}=0$, and
        $N^{-1}L_vh_{ab}=0$. Thus, if we require continuity of the matter
        variables through a surface of signature change, we found, as expected,
        that the second fundamental form must vanish correspondingly. Notice
        that this result follows  without requiring the a priori continuity of
        the 4-metric or the continuity of the second fundamental form.
        Actually, it is precisely the continuity of the matter variables which
        forces such a result. It is not in the geometry, and as argued in the
        previous paragraph, there is no a priori need to assume that for
        $\lambda=0$ the matter variables are continuous, or satisfy energy
        conditions reminiscent of the Lorentzian regime.\\

        In general, without imposing any continuity or sign restriction on
        $\hat{\mu}$, equation (37) must be considered as a constraint on the
        Lorentzian rate of volume expansion $k$. In other words, the above
        compatibility condition between the Hamiltonian constraints sets an
        origin for the {\it extrinsic time} $k$ which parametrizes the time
        evolution in the Lorentzian region. Geometrically speaking, this condi-
        tion is simply selecting the hypersurface where the signature change
        can occur [5,6]. \\

        One could use an approach even more closely tuned to the spirit of the
        initial value problem by using as dynamical variables the conformal
        part of the 3-metric, and the scaled divergence-free trace-free part
        of the second fundamental form. However this would complicate the
        equations without throwing much light on the basic issues we are
        addressing. We have therefore avoided these complications here,
        although this more detailed analysis shows signs of raising
        interesting questions.

        \section{Relation to other approaches}
        It is essential to our approach that the 3-metric is continuous
        through the change of signature. Others have emphasized [7,8,10] their
        belief in the importance of using coordinate systems where {\it all}
        the covariant components of the metric are continuous at a change of
        signature surface. We have not adopted this view, {\it inter alia}
        because then some of the contravariant metric components will diverge
        at the surface of change, leading {\it inter alia} to the divergence
        of various Christoffel terms; so the appearance of continuity is
        somewhat misleading. \\

             What we do believe is important is that the kinematics should be
        well-behaved there; this means we demand a well behaved shift and
        lapse, which determine the 4-dimensional metric structure. In
        particular the lapse should not go to zero because if it does then one
        halts the evolution in the coordinate system thereby defined. This
        means in turn that while the 3-metric components and their first
        `time' derivative can always be chosen continuous up to a
        diffeomorphism, if the lapse is regular then the 4-dimensional metric
        tensor components associated will have a discontinuous component (the
        time-time component, which is not dynamical). \\

        In our geometric approach, there is no need to assume {\it a priori}
        that the 4-dimensional metric is continuous, because we have shown
        that one can match the Lorentzian and Riemannian spacetimes without
        making such an assumption, by having a perfectly well behaved
        kinematical description (the lapse and shift are well-behaved in our
        approach). The kinematics through a signature change surface should as
        far as possible be free from particular coordinate choices, and one
        should be free to choose the kinematical data (the lapse and shift) as
        desired, not forced to make them go to zero.\\

        The approach of [7] is based on a different view:
        emphasizing more the role of the full space-time metric than the view
        used here. It also  assumes additional differentiability for the
        solutions, and is therefore more restrictive than the view adopted
        here; it is not surprising that the results obtained are more
        restrictive than if one does not impose these extra conditions.
        However that view also implies the lapse function goes to zero as one
        approaches the change surface. This `collapse of the lapse' may be
        expected to cause problems for the dynamics [18]. \\

        It will be clear from the above that the generic situation does not
        require a vanishing of the second fundamental form at the surface of
        change, which is required for example in both the distributional [7]
        and the Hartle-Hawking approach [19] (which uses a complex time
        variable). Our hope is that the present geometrical analysis of the
        classical case will be of help in understanding the full generality of
        what may be possible in the quantum case, through first clarifying the
        full generality of the analogous classical situation. \\ \\

        We thank the MURST (Italy) for support. We would like to thank
        Charles Hellaby for many useful criticisms and comments that
        greatly improved the presentation of the paper.\\ \\

        {\bf References}\\

        [1] S W Hawking, in {\it Astrophysical Cosmology}, Ed. H A Br\"{u}ck,
        G V Coyne and M S Longair (Pontifica Academia Scientarium, Vatican
        City, 1982), pp. 563-574.\\

        [2] J B Hartle and S W Hawking, {\it Phys Rev} {\bf D28}, 2960
        (1983).\\

        [3] S W Hawking, {\it Nucl Phys} {\bf B239}, 257 (1984).\\

        [4] T Dray, C Manogue and R Tucker, {\it Gen Rel Grav} {\bf 23}, 967
        (1991).\\

        [5] G F R Ellis, A Sumeruk, D Coule and C Hellaby, {\it Class Qu Grav}
        {\bf 9}, 1535 (1992).\\

        [6] G F R Ellis, {\it Gen Rel Grav} {\bf 24}, 1047 (1992).\\

        [7] S A Hayward, {\it Class Qu Grav} {\bf 9}, 1851 (1992).\\

        [8] S A Hayward, {\it Class Qu Grav} {\bf 10}, L7 (1993).\\

        [9] T Dray, C Manogue and R Tucker, {\it Phys Rev} {\bf D48}, 2587
        (1993). \\

        [10] M Kossowski and M Kriele, {\it Class Qu Grav} {\bf 10}, 2363
        (1993). \\

        [11] G F R Ellis and K Piotrkowska. To appear, {\it Proc Les
        Journ\'{e}es Relativistes}, Brussels, 1993. \\

        [12] C Hellaby and T Dray, `Failure of Standard Conservation Laws
        at a Classical Change of Signature'. To appear, {\it Phys Rev}
        {\bf D}, (1994).\\

        [13] W Israel, {\it Nuovo Cim} {\bf 44B} 1, {\bf 48B}, 463 (1966).\\

        [14] S W Hawking and G F R Ellis, {\it The Large Scale Structure of
        Space Time}. (Cambridge University Press, Cambridge, 1973).\\

        [15] A Fischer and J Marsden. In {\it General Relativity: An Einstein
        Centenary Survey}. Ed S W Hawking and W Israel (Cambridge University
        Press, Cambridge, 1979), pp. 138-202. \\

        [16] J Isenberg and J Nester. In {\it General Relativity and
        Gravitation. Vol I}, Ed A Held. (Plenum Press, 1980), p 73. \\

        [17] D M DeTurck, {\it J Diff Geom}, {\bf 18}, 157-162 (1983)
        (improved version); see also D M DeTurck, {\it Bull.
        Amer.Math.Soc}, {\bf 3}, 701-704 (1980).\\

        [18] J York, `Kinematics and Dynamics of General Relativity'. In {\it
        Sources of Gravitational Radiation}, ed. L. Smarr (Cambridge
        University Press, 1979), see pp. 103--110.\\

        [19] G W Gibbons and S W Hartle, {\it Phys Rev} {\bf D42}, 2458
        (1990).

        \end{document}